\documentclass{article}[15pt]

\usepackage{amsmath,epsfig,color,calc,rotating}
\usepackage{graphics}
\usepackage{wrapfig}
\usepackage{ulem}

\usepackage{natbib}
\bibpunct{(}{)}{;}{a}{}{,}

\begin{document}

\title{
Analyses of Baby Name Popularity Distribution in  \\
U.S. for the Last 131 Years
\vspace{0.2in}
\author{
Wentian Li \\
{\small \sl  The Robert S. Boas Center for Genomics and Human Genetics,
The Feinstein Institute for Medical Research} \\
{\small \sl North Shore LIJ Health System,
Manhasset, 350 Community Drive, NY 11030, USA.}\\
}
\date{}
}
\maketitle  
\markboth{\sl Li}{\sl Li}

We examine the complete dataset of baby name popularity
collected by U.S. Social Security Administration for
the last 131 years (1880-2010). The ranked baby name
popularity can be fitted empirically by a piecewise
function consisting of Beta function for the high-ranking
names and power-law function for low-ranking names,
but not power-law (Zipf's law) or Beta function by itself. 

\large

\section{Introduction}

\indent

Zipf's law describes a class of ranked distributions, in
which the ranked quantity $y$ falls off with rank $r$ by
$y_r = C/r^\alpha$ ($\alpha \approx 1$). This ``law" was
originally observed in the word usage in human languages \citep{zipf}
where $y_r$ is the number of times (number of ``token")
the rank-$r$ word (a ``type") appears in a language text.
But Zipf's law also fit many other datasets, such as 
city population \citep{brakman}, company size \citep{saichev},
internet traffic \citep{rowlands}, and many others \citep{wli-glott}.

When a ranked quantity deviates from a perfect Zipf's law,
it is often dismissed as a ``finite size effect", i.e.,
fluctuation in the ranked quantity for low-ranking entities
caused by small sample size. The assumption is that the fluctuation
will be reduced in a larger sample size, and the Zipf's 
law would be preserved. Unsatisfied with this
explanation, Cocho and his colleagues proposed a new class
of rank function similar in form to the Beta probability
distribution that curves in a log(rank)-log(quantity) plot
\citep{germi07,germi08,plos1}. This class of Beta rank functions
has been proven to be an excellent fitting function of
many datasets ranging from number of collaborators
in a social network, number of citations per article
and per journal \citep{germi07,campanario}, to musical score and alphabet
frequencies \citep{wli-entropy,roberto,wli-jql,petersen,wli-phya}.

The success of the Beta rank function may lead to a belief that
we may have a function universal enough to fit all ranked
datasets. Indeed, there have been works to prove a universal
mechanism behind the Beta rank function \citep{beltran},
or to relate the Beta rank function to a probability distribution \citep{sarabia}.
It was also an expectation that it may fit the baby name popularity 
data in U.S., to be analyzed in this paper, which is well documented by
the U.S. Social Security Administration since 1880.
If the number of babies with the same first name
(born in the same year and with the given gender) are ranked
by popularity ($\{ y_r \}$ for rank $r=1,2,\cdots$),
will the $y_r \sim r$ relationship be a power-law function (thus as
another example of the Zipf's law)? Or will it be a Beta ranked function?

\begin{figure}[th]
\begin{center}
  \begin{turn}{-90}
   \epsfig{file=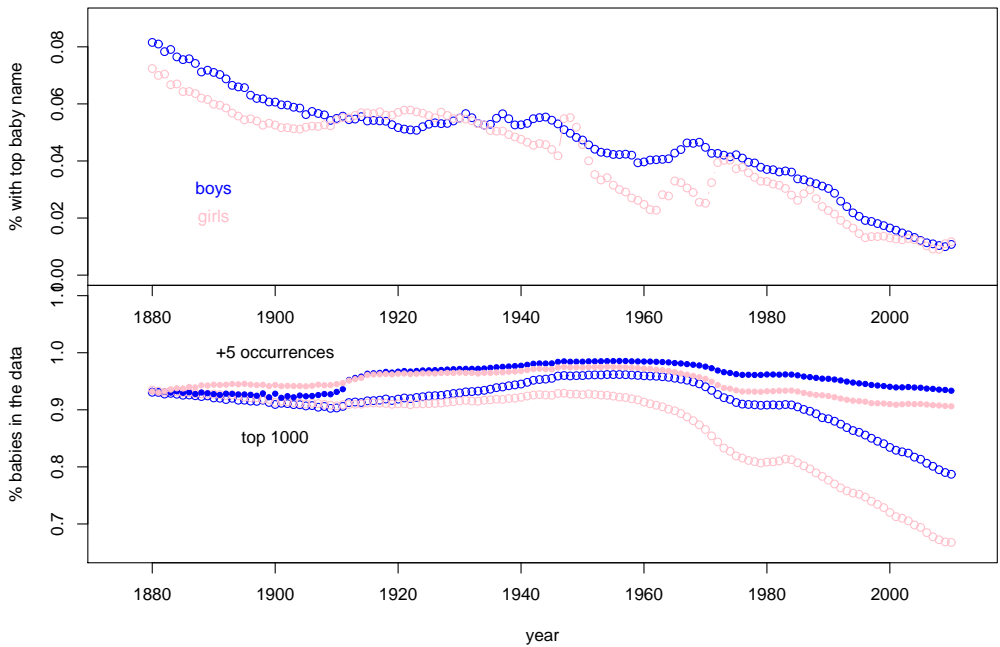, width=8.5cm}
  \end{turn}
\end{center}
\caption{
\label{fig1}
Top: The most popular boy (blue) and girl (pink) names as 
the percent of total births; Bottom: The numer of samples 
included in the datasets (top 1000 and +5 occurrences) 
as the percentage of total births.
}
\end{figure}

\section{The temporal trend of baby name popularity data}

\indent

The baby name popularity data in U.S. since 1880 is available
from the U.S. Social Security Administration (SSA) site: 
{\sl http://www.ssa.gov/oact/babynames/}. The boy and girl's
names are listed separately. Two versions of the data can be
downloaded from the website:  the web form version for top 1000 
(gender-specific) names, and the flatfile for all names with 5
or more occurrences. Baby names with less than 5 occurrences
are not given to the public to ``safeguard privacy". Since 
both data are not 100\% complete, we check how much of 
all births are included in the data by using the information 
of the top-1 baby name. The most popular baby name as a
percentage of all births is plotted in Fig.\ref{fig1} (top).
The number of all births (not directly provided by the
SSA site)  can be derived:  $n_{total}= n_{r=1}/p_{r=1}$.
Fig.\ref{fig1} (bottom) shows the percentage of births included
in the two datasets (top 1000 and +5 occurrences):
$p_{top1000}=\sum_{r=1}^{1000} n_{r}/n_{total}$,
$p_{+5}= \sum_{n_r \ge 5} n_{r}/n_{total}$

The top baby name was more dominant in early years than in recent
years (the top boy names were John (1880-1923), Robert (1924-1939, 1953),
James (1940-1952), Michael (1954-1959, 1961-1998), David (1960), 
Jacob (1999-2010), and the top girl names were
Mary (1880-1946, 1953-1961), Linda (1947-1952), Lisa (1962-1969),
Jennifer (1970-1984), Jessica (1985-1990, 1993-1995),
Ashley (1991-1992), Emily (1996-2007), Emma (2008), Isabella (2009-2010)).
However, to characterize the uneven naming of babies, using
the statistics from the top name alone is not enough. 

In economics, how the riches dominate over the poors in
the wealth distribution can be measured by the Gini index: 
if the riches have all the wealth,
the Gini index is 1; if there is no economic inequality, the
Gini index is 0 (see, e.g., \citep{cowell}). 
A formula for Gini index from the ranked data 
$\{ y_r \}$ is $G= R^{-1} (R+1- 2 \sum_{r=1}^R ry_r/\sum_{r=1}^R y_r)$
where $R=max(r)$ is the maximum rank (number of entities)
(implemented in, e.g., {\sl R} package {\sl reldist} \citep{handcock}).

\begin{figure}[th]
\begin{center}
  \begin{turn}{-90}
   \epsfig{file=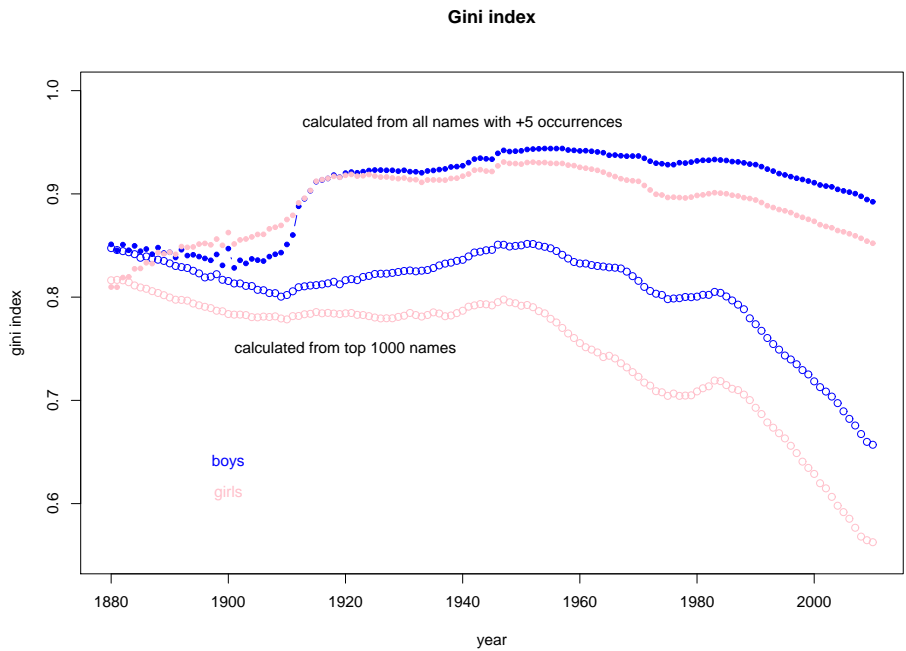, width=8.5cm}
  \end{turn}
\end{center}
\caption{
\label{fig2}
Gini index calculated for two datasets (top 1000 and +5 occurrences)
for boys (blue) and girls (pink)  separately.
}
\end{figure}

Fig.\ref{fig2} shows the Gini index calculated from both the top
1000 data and +5 occurrence data for the last 131 years.
There are following observations: First of all, using only the 
top 1000 names will underestimate the Gini index, and the degree 
of underestimation is more severe with larger sample sizes. It is because the
baby names not included in the data contribute more to
the inequality of name popularity. With more babies born in recent
years, top 1000 names cover less (Fig.\ref{fig1} bottom) of all births. 

Secondly,  there are more diversities (inequalities), 
generally speaking, in girl names than in boy names. 
A  similar conclusion was reached in \citep{barry} by a
regional data -- the state of Pennsylvania --  from comparing 
baby names in only two years, 1960 and 1990. This can partially 
be seen from Fig.\ref{fig1} as for most years, the top girl 
names tend to take a smaller percentage of all
girl births than the corresponding top boy names. 

Thirdly, there is a weak trend of lower Gini index in recent 
(e.g. 30) years, or more diverse name givings.  A similar 
conclusion was also reached in \citep{twenge}. This can be 
caused by many factors, such as more diverse immigrant groups 
in recent years. It is not clear how the increase of total
number of names may contribute to this trend.

\begin{figure}[th]
\begin{center}
  \begin{turn}{-90}
   \epsfig{file=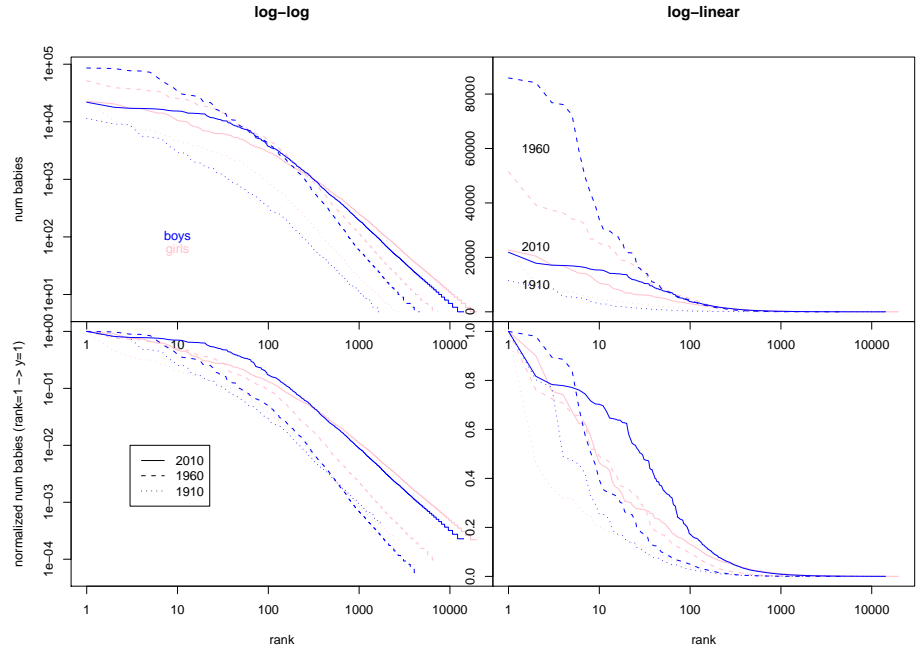, width=8.5cm}
  \end{turn}
\end{center}
\caption{
\label{fig3}
Four versions of the ranked baby name popularity data
for the years 1910, 1960, and 2010 (boys (blue) and girls
(pink) are ranked separately): (upper left) log-log;
(upper right) log-linear; (lower left) log-log
where the $y$ is normalized by the number of the most popular
baby name (i.e., rank-1 name has $y$ value equal to 1);
(lower right) log-linear where $y$ is normalized.
}
\end{figure}

\section{Fitting ranked baby name popularity}

\indent

\begin{figure}[th]
\begin{center}
  \begin{turn}{-90}
   \epsfig{file=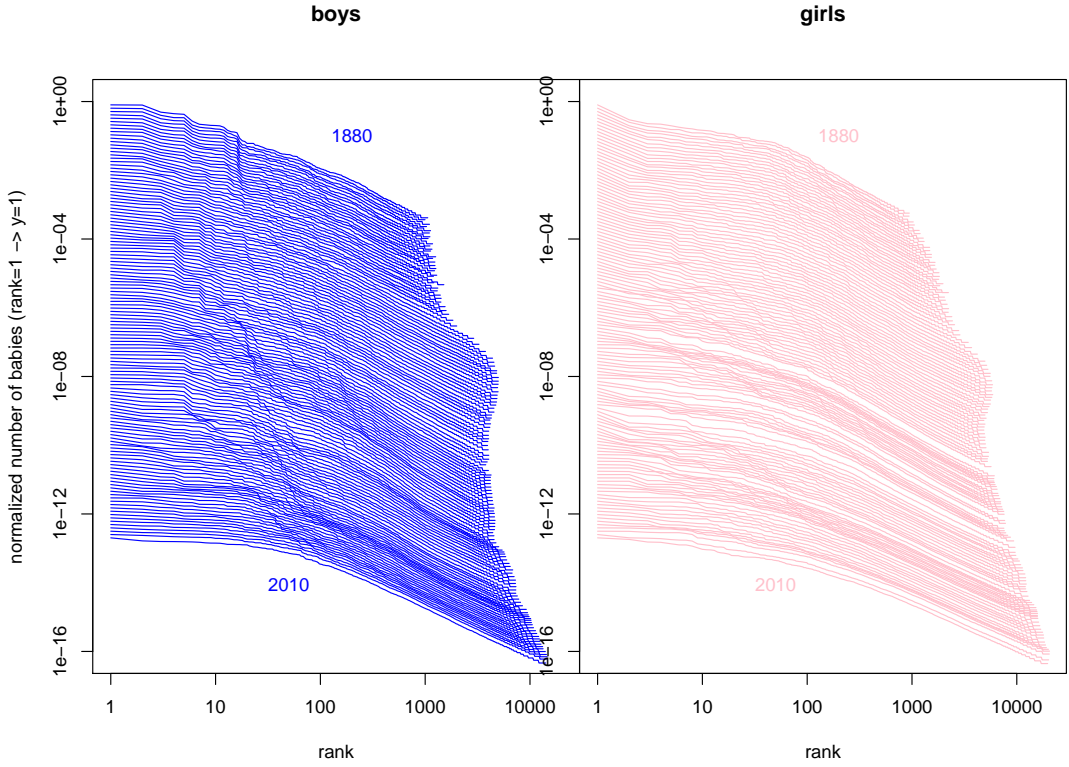, width=10cm}
  \end{turn}
\end{center}
\caption{
\label{fig4}
Ranked normalized number of baby names (rank-1 name,
or the most popular name,  has $y$ value equal to 1) 
for each year from 1880 to 2010 (131 lines).  Boys
(blue)  and girls (pink)  names are ranked separately.
Plots from different years are shift downward to 
improve visibility.
}
\end{figure}

To find a better representation of the ranked distribution of baby
name popularity, we show 4 different versions of the ranked
distribution for the (+5 occurrence) data in years 1910, 1960, and 2010 
in Fig.\ref{fig3}. In principle, there are 16= 4 $\times$ 4
versions of plotting, first 4 for linear($x$)-linear($y$), linear-log, 
log-linear, log-log, and the other 4 for original($x$)-original($y$), 
normalized-original, original-normalized, and normalized-normalized. 
By normalizing, we mean divided by the maximum value. We 
discard the linear-linear and linear-log (because high-ranking 
names are not highlighted enough in these versions),
we also discard the original-normalized and normalized-normalized
(because the total number of baby names information is not available).
For the remaining 4 versions in Fig.\ref{fig3}, the log-log version
seems to be the better representation as rank distributions from
different years are closer to each other.  The ranked normalized 
baby name popularities in log-log scale for every year from 
1880 to 2010 are shown in Fig.\ref{fig4} (lines are shifted downward 
to improve visibility).

Each year's ranked normalized baby name popularity data in
Fig.\ref{fig4} is fitted  by three functions, power-law function,
Beta rank function, and a two-piece function:
\begin{eqnarray}
\label{eq-3fun}
\mbox{power-law:} \hspace{0.1in}
\log(y_r/y_{r=1}) &=& C - a \log(r) \nonumber \\
\mbox{Beta:} \hspace{0.1in}
\log(y_r/y_{r=1}) &=& C - a \log(r) + b \log (R+1-r) \nonumber \\
\mbox{two-piece:} \hspace{0.1in}
\log(y_r/y_{r=1}) &=& 
\left\{
\begin{array}{lc}
C - a \log(r) + b \log (R+1-r) & \mbox{ $r< r_0$} \\
C - a \log(r) & \mbox{ $r \ge r_0$} 
\end{array}
\right.
\end{eqnarray}
The first is the power-law function (Zipf's law),
the second is the Beta function \citep{plos1}, where
$R=\max(r)$ is the maximum rank in the +5 occurrence data
(the total number of all baby names, including
those with less than 5 occurrences, is unavailable),
and the last function is a two-piece function with
Beta function for high-ranking and power-law
function for low-ranking regimes.

The two-piece function is motivated by inspecting the
ranked plots in Fig.\ref{fig4}: the high-ranking points
seem to follow a straight line, whereas the low-ranking
points are curved in the log-log plot. This choice 
is also motivated by the fact that the fitting
of Beta function in the linear-linear scale by the 
non-linear least-square regression \citep{nls,wli-jql,wli-phya}
does not converge for these datasets (results not shown),
indicating that Beta function does not fit the data well
in the full rank range.
 
The segmentation point $r_0$ is not known beforehand.
We determine its value by minimizing the overall
sum-of-squared-error (SSE):
\begin{eqnarray}
\label{eq-sse}
SSE(r') &= &\sum_{r=1}^{r'} 
\left( \log \frac{y_r}{max(y)}- C+a\log(r)-b \log(R+1-r)
\right)^2
 \nonumber \\
&& + \sum_{r=r'+1}^R 
\left( \log \frac{y_r}{max(y)} - C+a\log(r) \right)^2
 \nonumber \\
s_0 &=& \arg\max_{r'} SSE(r')
\end{eqnarray}
To save computing time, we limit the range of choice of
$r_0$ between $r=20$ and $r=200$.

\begin{figure}[th]
\begin{center}
  \begin{turn}{-90}
   \epsfig{file=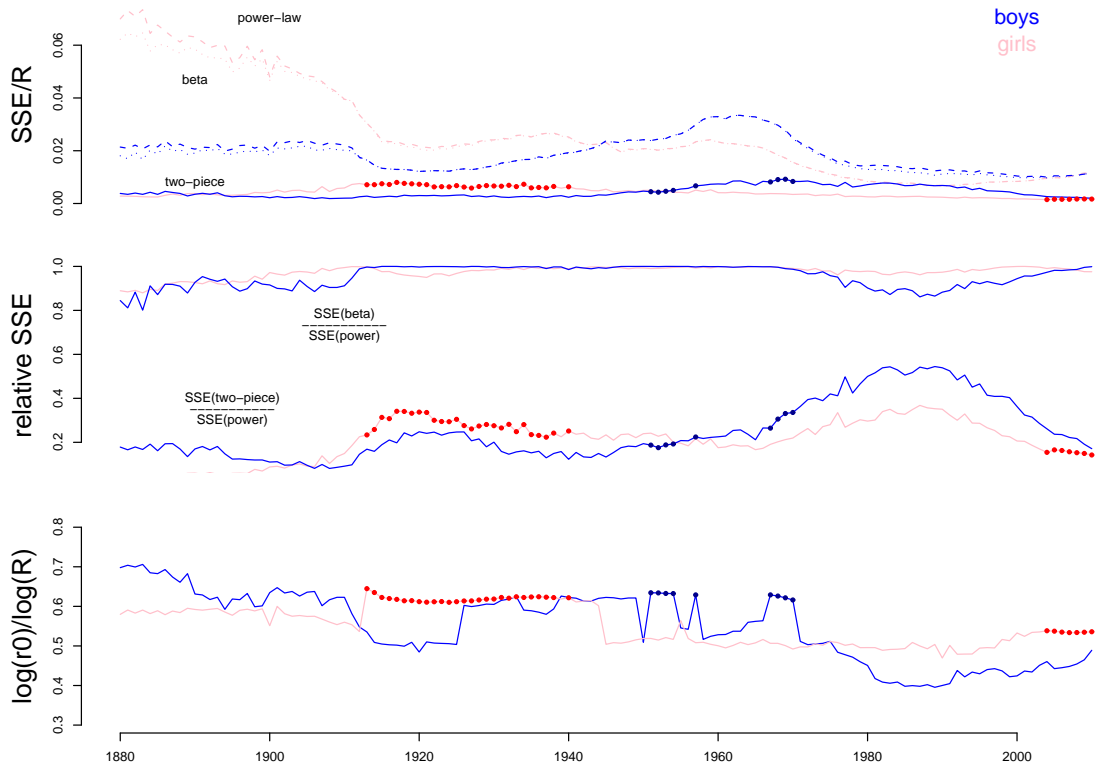, width=10cm}
  \end{turn}
\end{center}
\caption{
\label{fig5}
Results for boy names are marked by blue, and those for girl 
names by pink.
(1) Sum of square errors (SSE) (Eq.(\ref{eq-sse}) divided 
by the maximum rank ($R$) for power-law, Beta, and 
two-piece function in Eq.(\ref{eq-3fun});
(2) SSE of Beta and two-piece function normalized by the
SSE of the power-law function;
(3) relative position of the point which separates the
Beta and power-law regimes in the two-piece function,
in logarithmic scale: $\log(r_0)/\log(R)$.
}
\end{figure}

\begin{figure}[th]
\begin{center}
  \begin{turn}{-90}
   \epsfig{file=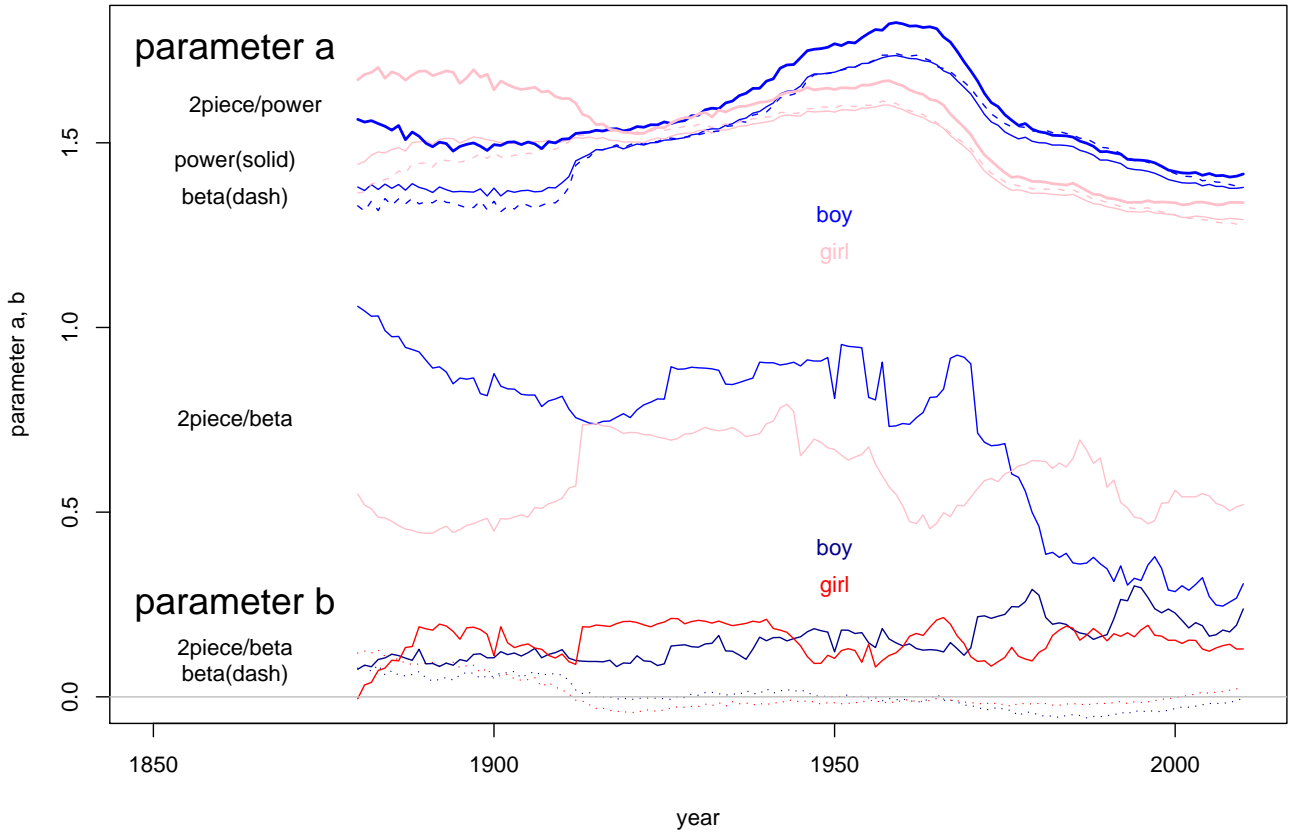, width=11cm}
  \end{turn}
\end{center}
\caption{
\label{fig6}
Results for boy names are marked by blue or darkblue, 
and those for girl names by pink or red.
See Eq.(\ref{eq-3fun}) for definitions.
The $a$ values for power-law, Beta function, the power-law 
regime of the two-piece function, and the Beta regime of 
the two-piece function, the $b$ values for Beta function, 
the Beta regime of the two-piece function.
}
\end{figure}

Fig.\ref{fig5} and Fig.\ref{fig6} show all fitting results 
by these three functions for the baby popularity data 
of the last 131 years. The top plot is the SSE per point 
(i.e., SSE in Eq.(\ref{eq-sse}) divided by the maximin rank)
for the power-law function, Beta function, and two-piece
function. The data for boys is in blue color, that for
girls in pink. The Beta function reduces the SSE
as compared to the power-law function, but the reduction
is not large. On the other hand, the two-piece function
fits the data much better. For the two-piece function,
if the $r_0$ is chosen at the boundary of our selected
range (e.g. $r_0=200$), it indicates that SSE may be
reduced even further if we remove the constraint.
These datasets are marked by red/darkblue dots in Fig.\ref{fig5}.

The second plot in Fig.\ref{fig5} is the relative SSE:
SSE for Beta function, and SSE for the two-piece function,
over that for the power-law function. The ratio is only
slighly less than 1 for the Beta function, but is as low as
0.2-0.3 for two-piece functions.

The third plot in Fig.\ref{fig5} is the relative position of the
segmentation point in the logarithmic scale: 
$\log(r_0)/\log(R)$. If this value is 0.5, then
Beta and power-law function covers half of the
log-rank range in the two-piece function.

Fig.\ref{fig6} shows the fitted parameter values for 
$a$ in the power-law function, in Beta-function, in 
both parts in the two-piece function.
The value of $a$ is mostly between 1 and 2, but for the
Beta function in the two-piece situation, is mostly
less than 1. The $b$ value in Beta function fluctuates
around 0, but for the Beta function in the two-piece
situation becomes much larger (median value is 0.14-0.16).

All these results point to the conclusion that our two-piece
function, with Beta function for the low-ranking data
and power-law function for the high-ranking data,
fits the data much better than both power-law function
and Beta function by themselves. Graphically speaking,
it is another way to state that in log-log plot,
the ranked baby name popularity in a typical year
curves down, then followed by  a straight line.

\section{Discussion}

\indent

The universality of power-law distribution in nature as
well as human-related data is heatedly debated \citep{clauset}.
To require a power-law to be true for all ranges is
clearly a very strong constraint, and not many datasets
can pass the test. The introduction of Beta rank function 
\citep{plos1} is to relax the strong requirement and
increase the number of datasets to be well fitted.
The main contribution of this paper is that even Beta
rank function may fail to fit some dataset well, such
as the baby name popularity data.

Our choice of using Beta function for high-ranking
names and power-law function for low-ranking names
is empirically based on examining the raw data in log-log
plot, but it has interesting parallels in other fields.
It has been observed in linguistic data that Beta rank
function tends to fit the data very well when the maximum
rank is limited, for example, for the ranked
letter frequency distribution \citep{wli-jql}.
On the other hand, it may not be the case when there is 
no limit on the maximum rank, such as the word frequency 
distribution, where the mechanism for data generation 
is described by the ``large number of rare events" model \citep{baayen}. 

We can imagine a situation in which quantitative
law governing the popular names (fewer and limited)
distribution being distinct from that for rare names
(numerous and unlimited). A similar idea in separating
high-ranking and low-ranking events in another application
(polymorphous Chinese syllables versus regular syllables)
was discussed in \citep{wli-jql2}. 

As in the case of any empirical fitting of data,
other empirical alternatives are possible. If particular,
with only 100 or less (median of the first regime
in the two-piece function) points to fit, many other
functions other than Beta rank function may also be
used. If the heterogeneity between high- and low-ranking
names does indeed exist, then it is a challenge to find a single
functional form which could fit the ranked baby popularity
data in its whole rank range.

\section*{Acknowledgements}
\indent

I would like to thank Pedro Miramontes and Jan Freudenberg for 
comments and the Robert S Boas Center for Genomics and Human Genetics
for support.

\end{document}